\documentstyle[12pt]{article}
\begin{document}
\vskip 4 cm
\begin{center}
\Large{\bf ON THE ORIGIN OF THE UNIVERSE}
\end{center}
\vskip 3 cm
\begin{center}
{\bf AFSAR ABBAS} \\
Institute of Physics, Bhubaneswar-751005, India \\
(e-mail : afsar@iopb.res.in)
\end{center}
\vskip 20 mm  
\begin{centerline}
{\bf Abstract }
\end{centerline}
\vskip 3 mm

It has been proven recently that the Standard Model of particle
physics has electric charge quantization built into it. 
It has also been shown by the author that there was no electric 
charge in the early universe. Further it is shown here that the
restoration of the full Standard Model
symmetry ( as in the Early Universe ) leads to the result that `time',
`light', along with it's velocity c and the theory of relativity,
all lose any physical meaning. The physical Universe as we know it,
with its space-time structure, disappears in this phase transition.
Hence it is hypothesized here that the Universe came into existence when
the Standard Model symmetry $ SU(3)_C \otimes SU(2)_L \otimes U(1)_Y $
was spontaneously broken to $ SU(3)_C \otimes U(1)_{em} $.
This does not require any spurious extensions of the Standard Model
and in a simple and consistent manner explains the origin of
the Universe within the framework of the Standard Model itself. 

\newpage

In the currently popular Standard Model of cosmology the Universe is
believed to have originated in a Big Bang. Thereafter it started
expanding and cooling. Right in the initial stages, it is believed that
models like superstring theories, supergravity, and GUTs etc would be 
applicable. When the Universe cooled to 100 GeV or so, the Standard
Model of particle physics (SM) symmetry of $ SU(3)_C \otimes
SU(2)_L \otimes U(1)_Y $ was spontaneously broken to
$ SU(3)_C \otimes U(1)_{em} $. Thereafter the Universe went through
quark-gluon to hadron phase transition etc. Belief in the correctness of
this model is so prevalent that many are confident that we already have 
inkling of the Theory of Everything.

However it has recently been shown by the author that models like
the superstring theories, supergravity,
and GUTs etc have a fatal flaw
in them in as much as they are inconsistent with the SM [1,2].
As the SM has been well verified experimentally and is so far the best
model of particle physics, this inconsistency rules them out. To
understand this point and to obtain further understanding let us start
by summarizing the arguments [1,2].

It has been shown by the author [1] that in SM 
$ SU(N_{C}) \otimes SU(2)_{L} \times U(1)_{Y} $
for $ N_{C} = 3 $
spontaneous symmetry breaking by a Higgs of weak hypercharge
$ Y_{ \phi } $
and general isospin T where $ T_{ 3 }^{ \phi } $ component 
develops the vacuum
expectation value $ < \phi >_{0} $, fixes `h' in the electric charge
definition $ Q = T_{ 3 } + h Y $ to give

\begin{equation}
Q = T_{ 3 } - \frac{ T_{ 3 }^{ \phi } }{ Y_{ \phi } } \, Y
\end{equation}

where Y is the hypercharge for doublets and singlets for
a single generation.
For each generation renormalizability through triangular anomaly
cancellation and the requirement of the identity of L- and R-handed charges
in $ U(1)_{ em } $ one finds that all unknown hypercharges are
proportional to 
$ \frac{ Y_{\phi} }{ T_{ 3 }^{ \phi } } $. Hence correct charges (for 
$ N_{C} = 3 $ ) fall through as below [1]

\begin{eqnarray}
Q(u) = \frac{1}{2} ( 1 + \frac{1}{N_{C}} ) \\
Q(d) = \frac{1}{2} ( -1 + \frac{1}{N_{C}} ) \\
Q(e) = -1 \\
Q(\nu) = 0
\end{eqnarray}

Note that the expression for Q in (1) arose due to spontaneous symmetry
breaking of $ SU(N_{C}) \otimes SU(2)_{L} \otimes U(1)_{Y} $
(for $ N_{C} = 3 $ ) to 
$ SU(N_{C}) \otimes U(1)_{em} $ through the medium of a Higgs with
arbitrary isospin T and hypercharge $ Y_{\phi} $. What happens when at
higher temperature, as for example found in the early Universe, the 
$ SU(N_{C}) \otimes SU(2)_{L} \otimes U(1)_{Y} $ symmetry is restored ?
Then the parameter `h' in the electric charge definition remains
undetermined. Note that `h' was fixed as in (1) due to spontaneous
symmetry breaking through Higgs. Without it `h' remains unknown. Thus the
electric charge loses all physical meaning above the EW phase transition
stage and thus one concludes that as per the SM there was no electric
charge in the early Universe. 

Note that complete charge quantization in the canonical Standard Model
with a Higgs doublet had been demonstarted by the author earlier
[3,4,5,6]. Hence there too on
the restoration of the full symmetry the electric charge would disappear
as shown above for a Higgs with any arbitrary isospin and hypercharge. 
There [3,4,5,6] as well as here, `h' is not defined and hence the electric
charge is not defined. Thus when the
electro-weak symmetry is restored, irrespective of the Higgs isospin and
hypercharge the electric charge
disappears as a physical quantity. Hence there too we find that there was
no electric charge in the early universe.

It is a generic property of all models like the superstring theories,
supergravity and GUTs etc that they have
electric charges which are quantized. As these models are 
believed to be describing physics above the electro-weak symmetry breaking
stage, they are all inconsistent with the SM requirement that then there
was no electric charge above this scale.
Hence these are ruled out [1].

All this misunderstanding and confusion arose because 
until now we had not understood
the electro-weak phase transition correctly. 
Here we have shown that the
restortion of the full SM symmetry leads to the result that there
is no electric charge above it and also that there was no photon.
What else?

The above result shows that above the electro-weak  phase transition there
was no electromagnetism. Maxwells equations of electrodynamics show that
light is an electromagnetic phenomenon. Hence above the electro-weak phase
transition there was no light and no Maxwells equations. And surprisingly
we are led to conclude that there was no velocity of light c as well. One
knows that the velocity of light c is given as
 
\begin{equation}
c^2 = { 1 \over { \epsilon_0 \mu_0} }
\end{equation}

where $ \epsilon_0  $ and $ \mu_0 $ are permittivity and permeability
of the free space. These electromagnetic properties disappear
along with the electric charge and hence the velocity of light also
disappears above the electro-weak phase transition.

The premise on which the theory of relativity is based is 
that c, the velocity of light is always the same, no matter from which
frame of reference it is measured. Relativity theory also asserts that
there is no absolute standard of motion in the Universe. Of
fundamental significance is the invariant interval

\begin{equation}
s^2 = { ( c t ) }^2 - x^2 - y^2 - z^2
\end{equation}

Here the constant c provides us with a means of defining time in terms of
spatial separation and vice versa through l = c t. This enables one to
visualize time as the fourth dimension. Hence time gets defined due to
the constant c. Therefore when there were no c as in the early universe 
,there was no time as well. Hence above the electro-weak breaking scale 
there was no time. As the special theory of relativity depends upon c and
time, above the electro-weak breaking scale
there was no special theory of relativity. As the General Theory of
relativity also requires the concept of a physical space-time, it
collapses too above the electro-weak breaking scale. Thus there was no
gravity too above this scale.
Hence the whole physical universe collapses above this scale. 

It is hypothesized here that the Universe came into existence
when the electro-weak symmetry was broken spontaneously.
Before it there was no electric charge, 
no 'time', no 'light', no maximum and constant
velocity like 'c', no gravity or space-time structure on which objective
physical laws could exist.

Note that this new picture of the origin of the Universe arises naturally
within the best studied and the best verified 
model of particle physics. It does not require any
ad-hoc or arbitrary models or principles. One only has to do a consistent 
and careful analysis of the hitherto misunderstood electro-weak
spontaneous symmetry breaking. As should be
clear now, the Standard Model of the particle physics tells us as to how
the Universe was created ( as shown in this paper ), gives hints as to
what it was like before this
and also as to how it functions thereafter.
 
\vskip 4 cm
\begin{center}
{\bf\large REFERENCES }
\end{center}

\vskip 2 cm

1. Abbas A., 2000,
"Particles, Strings and Cosmology PASCOS99", Ed K Cheung,
J F Gunion and S Mrenna, World Scientific, Singapore ( 2000),
p. 123 - 126

2. Abbas A., July 1999,
{\it Physics Today }, p.81-82

3. Abbas A., 1990,
{\it Phys. Lett. }, {\bf B 238}, 344

4. Abbas A., 1990,
{\it J. Phys. }, {\bf G 16 }, L163

5. Abbas A., 1992,
{\it Hadronic J. }, {\bf 15 }, 475

6. Abbas A., 1993,
{\it Nuovo Cimento }, {\bf 106 A }, 985 

\end{document}